\pdfoutput=1
\documentclass[10pt]{article}
\usepackage{amssymb,amsmath,latexsym}

\usepackage[sectionbib]{chapterbib}
\usepackage{geometry}

\usepackage[USenglish,english,american]{babel}

\usepackage[utf8]{inputenc}
\usepackage{graphicx}
\usepackage{multirow}
\usepackage{amsmath}
\usepackage{amssymb}
\usepackage{amsthm}
\usepackage{amsfonts}
\usepackage{tipa}
\usepackage{textcomp}
\usepackage{array}
\usepackage{framed}
\usepackage{float}
\usepackage{sidecap}

\usepackage{fix-cm} 
\usepackage{palatino}
\usepackage[margin=1cm]{caption}
\usepackage{lettrine}

\usepackage[dvipsnames]{color}
\usepackage{listings}
\usepackage{cite}
\usepackage{ctable}
\definecolor{shadecolor}{gray}{0.9}
\usepackage[usenames,dvipsnames]{xcolor}
\usepackage{doi}
\usepackage{url}

\usepackage{authblk}

\title{More than one Author with different Affiliations}
\author[1,2]{Octavio Miramontes}
\author[3]{Og DeSouza}
\author[3,4]{Leticia Ribeiro Paiva}
\author[3]{Alessandra Marins}
\author[1]{Sirio Orozco}
\affil[1]{\small{Instituto de Fisica, UNAM, Mexico 04510 DF, Mexico}}
\affil[2]{\small{Centro de Ciencias de la Complejidad C3, UNAM, Mexico 04510 DF, Mexico}}
\affil[3]{Departmento de Entomologia, UFV, MG 36570-000, Brazil}
\affil[4]{Departamento de F\'isica, Universidade Federal de Ouro Preto, Ouro Preto, MG, Brazil}

\begin{document}
\title{L\'evy flights and self-similar exploratory behaviour of termite workers: beyond model fitting}
\maketitle
\abstract{
Animal movements have been related to optimal foraging strategies where self-similar trajectories are central. Most of the experimental studies done so far have focused mainly on fitting statistical models to data in order to test for movement patterns described by power-laws.
Here we show by analyzing over half a million movement displacements that isolated termite workers actually exhibit a range of very interesting dynamical properties --including L\'evy flights-- in their exploratory behaviour. Going beyond the current trend of statistical model fitting alone, our study analyses anomalous diffusion and structure functions to estimate values of the scaling exponents describing displacement statistics. We evince the fractal nature of the movement patterns and show how the scaling exponents describing termite space exploration intriguingly comply with  mathematical relations found in the physics of transport phenomena.   By doing this, we rescue a rich variety of physical and biological phenomenology that can be potentially important and meaningful for the study of complex animal behavior and, in particular, for the study of how patterns of exploratory behaviour of individual social insects may impact not only their feeding demands but also nestmate encounter patterns and, hence, their dynamics at the social scale.
}

\section{Introduction}

Subterranean termites are tropical social insects having an economic impact in the billions of dollars all over the world each year \cite{su1990economically}. For this reason, their patterns of movement have fomented both practical and theoretical studies in the past. Most of these studies, however, have concentrated in displacements at colony level and inside tunnels leaving the study at the individual level largely overlooked. Also, while animal movement in the context of resource acquisition has received a fair deal of attention \cite{sims2008scaling, viswanathan2008levy, bartumeus2008fractal, reynolds2009levy}, its links to individual interaction and --ultimately-- to sociality still need better focus. Providing that interindividual interactions lie at the very heart of sociality \cite{Miramontes.DeSouza.2008} and that interactivity depends on movement and space \cite{DeSouza.Miramontes.2004}, effective space exploration must play prominent role in social behaviour in general, and in termites in particular \cite{miramontes2014social}.

A common misconception about termite spatial orientation is that physical restrictions imposed by tunnel walls would guide effective space sweeping. While this could be true for termites foraging within tunnels underground, once inside their nest the scenario changes completely. There, galleries merge and split in unpredictable ways usually forming an entangled  set of paths and chambers whose layout does not seem to help orientation in most cases. Termites could, then, be guided by specific chemical cues and vibratory signals to effectively find nestmates. But would that be entirely dependent on external clues? Or, as sustained by a long forgotten hypothesis \cite{Jander.Daumer.1974}, would termites orient themselves concatenating external stimuli with an optimized search strategy, as with other animals \cite{boyer2006scale}?

Here we present evidence in favour of such a hypothesis, describing exploratory spatial behaviour in isolated termite workers kept in large containers, free from the constrained movements they experience within tunnels. In this way we were able to assess individual free exploratory behaviour in clueless environments and away from social interactions. We conclude that their searching patterns are compatible with scale-free strategies based on a fractal exploration of space and that these are key to the efficient flow of information between nestmates, thereby providing expressive hints on how self-organization underlies social cohesion \cite{Sumpter.2006} 

Our study includes the analysis of anomalous diffusion where the mean squared displacement (MSD) is estimated as a base to identify superdiffusive aspects of termites movements. Within the framework of anomalous diffusion theory \cite{klafter1987stochastic,ramos2004levy }, the scaling exponent of the MSD is related mathematically to a L\'evy probability distribution $P(l)\propto l^{-\mu}$ of the steps lengths $l$, with $1<\mu\leq3$. We evaluate these relationships and compare them versus the results that can be extracted from a Kolmogorov structure functions analysis. We progress into exhibiting that a termite walking is a $1/f$ noise process with power-laws in the correlation function indicating long-memory that is compatible with the fractal structures revealed by an Iterated Function System algorithm. Finally, we use maximum likelyhood estimation (MLE) to  show that waiting times with power-law scaling exponent values are also present in the termite exploratory behaviour with scaling exponents values compatible with those predicted by the physics of transport phenomena. 

\begin{figure}[!ht]
\centering
\includegraphics[width=100mm]{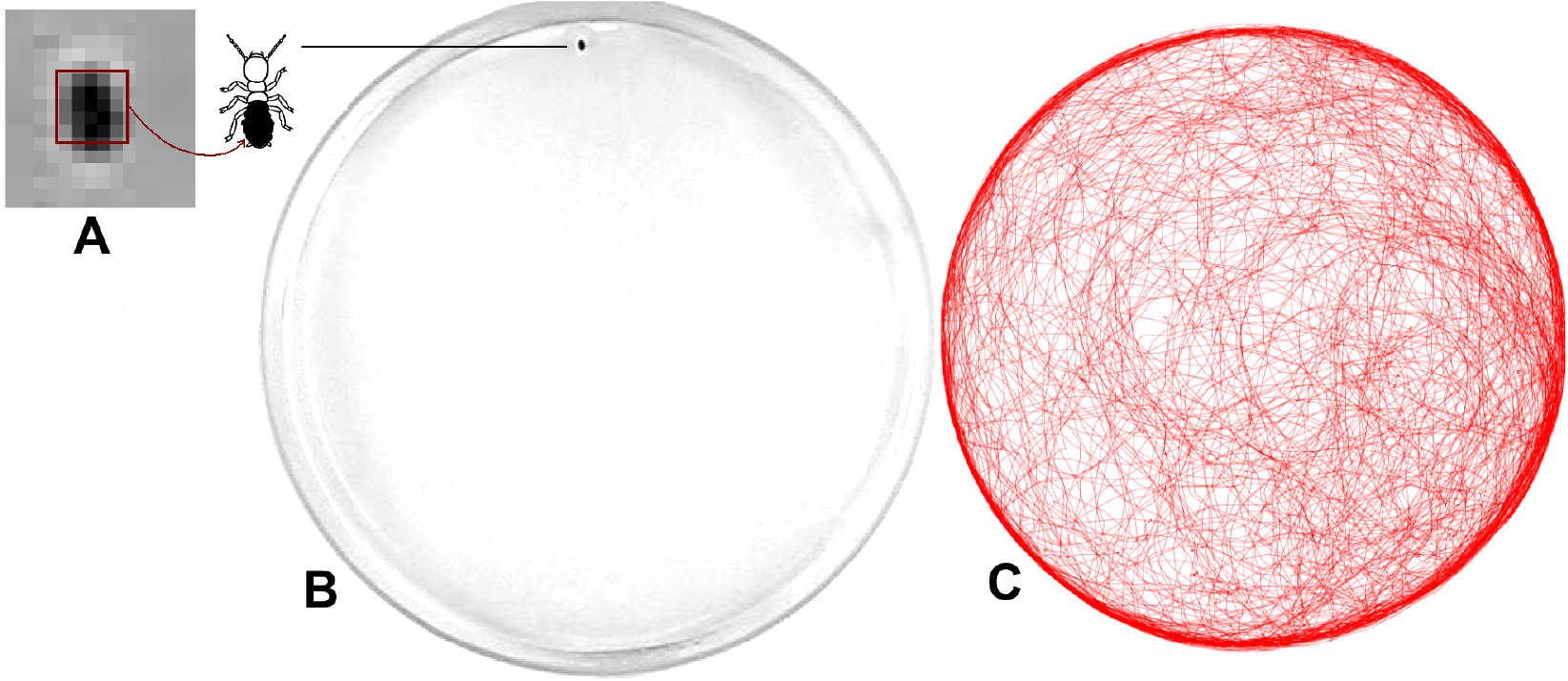}
\caption{{ \small{Picture of the observation set.} (A) One \emph{Cornitermes cumulans} worker with a painted abdomen was allocated into a circular glass arena (205mm inner dia.). When detected by the video recording system, the individual appears as an image measuring 5$\times$5 pixels representing 4.7 mm$^2$ aprox. In (B) the termite worker is the small black dot at the top of the circular area. A single total trajectory is drawn in (C) showing the typical entangled pattern of individual steps. This particular example contained 35,000 points sampled at 0.5 seconds intervals. Notice that most of the trajectory occurs near the arena border, however inner exploratory excursions are also frequent.}}
\label{fig:arena}
\end{figure}

\section*{Materials and Methods}

\subsection*{Experimental set-up}\hspace{2mm}\textit{Cornitermes cumulans} (Kollar) (Blattaria: Isoptera: Termitidae: Nasutitermitinae) workers were collected from a wild colony at the campus of the Federal University of Vi\c cosa, Minas Gerais, Brazil (20$^\circ$ 45' S, 42$^\circ$ 52' W), on 14 Dec 2010. \emph{Cornitermes} spp. are Neotropical termites occurring in several habitats, including forests, ``cerrados'' (Brazilian savannas), and man-modified habitats. No specific permissions were required for the locations or activities reported in this manuscript. The field studies did not involve endangered or protected species. The study did not involved human participants, specimens or tissue samples, or vertebrate animals, embryos or tissues.

\subsection*{Movement recordings}\hspace{2mm}Experimental arena  consisted of a glass Petri dish (\O{} = 205 mm) upside down over a sand-blasted flat glass. To allow evaluation of free-walking behaviour, no obstacle, nestmates or food was present in the arenas. Within the arena, a single termite worker was inserted and its movement was tracked from above, continuously for 5--6 hours, with a closed-circuit video camera (Panasonic WV-BP334 B\&W CCD) equipped with a macro lens (Fujinon YV5$\times$2.7 R4B-SA2L). To allow for full contrast between termites and the background, thereby permitting appropriate video recording, termites were painted with non-toxic water soluble dye \cite{Brunow.etal.2005}. We also performed experiments in containers of diameters 140 mm, 90 mm and 55 mm and found that the size did not significantly altered the results here reported. In total, half a million worth of termite displacements were recorded. We concentrate in reporting the data from the largest containers and choose four representative examples for our analysis and discussion.  

\begin{figure}[!h]
\centering
\includegraphics[width=70mm]{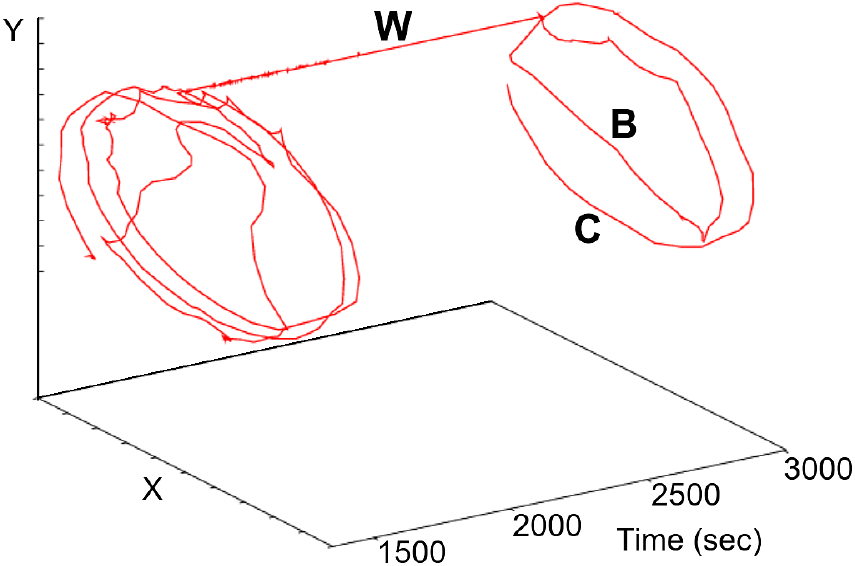}
\caption{{ \small{A termite walking trajectory segment as reconstructed in 3D, with a time axis added.} Three walking behaviours are visible. (C) is the individual walking termite following the circular geometry of the container border, (B) is a straight long nearly-ballistic displacement from one container side to the opposite, and (W) is a waiting time.}}
\label{fig:tra}
\end{figure}

Termite trajectories were captured and digitized at a sample rate of one point every 0.5 s with an automatic video-tracking software (EthovisionXT version 8.5.614, Noldus Information Technology). The paths were converted as a series of Cartesian coordinates coupled with respective time records, which allowed posterior calculations of spatial-temporal displacements of termites and their associated parameters, as shown in Fig. \ref{fig:arena}. Recordings included all three displacing behaviours presented by termites in the arena: (i) walking along the border of the arena, (ii) free moving, and (iii) static waiting-times (see Fig.\ref{fig:tra}). 

\begin{figure}[!h]
\centering
\includegraphics[width=90mm]{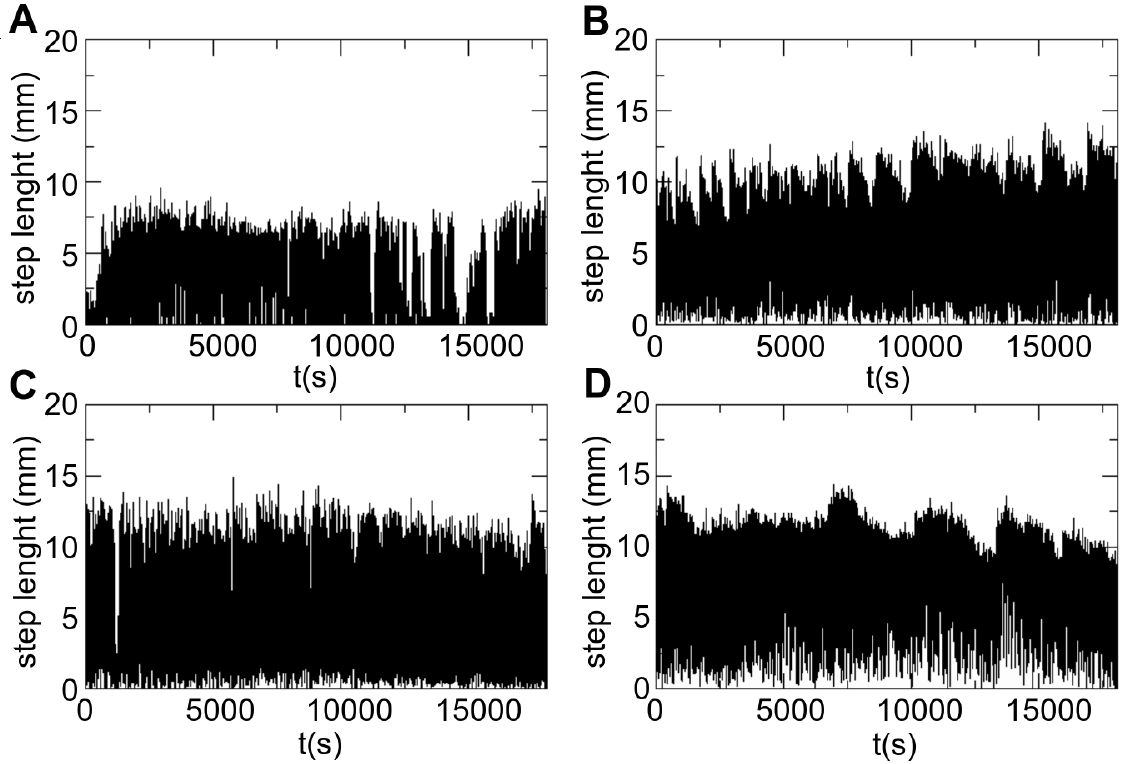}
\caption{{ \small{Examples of time-series containing traveled distances by four different \emph{Cornitermes cumulans} individuals.}  The time-series totaled 35,000 data points in (A) and 43,000 in (B, C and D), but a window of 18,000 points is shown for each. Sample rate was one point at every 0.5 seconds.}
\label{fig:distances}}
\end{figure}

\begin{figure}[!h]
\centering
\includegraphics[width=50mm]{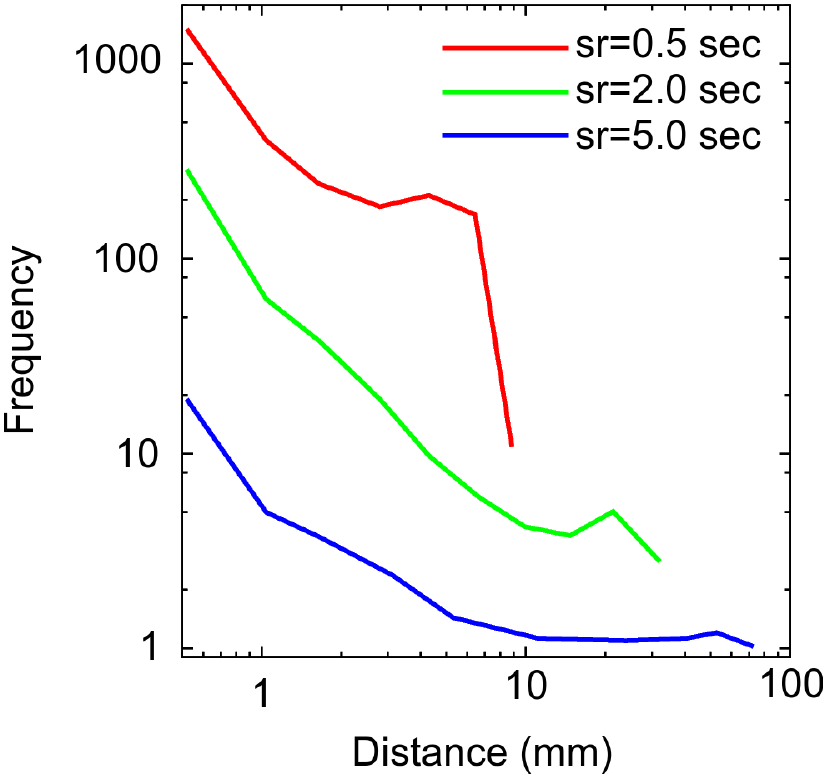}
\caption{{\small{Sampling rates.} Different sampling rates give very different results when the length of the steps are plotted as a histogram of their frequency (log-binned). In the plot, three different sampling rates (sr) were exemplified as the time series as captured by the video recording device at each 0.5, 2.0, and 5.0 sec. Note that a region resembling a power-law scaling is only obvious at 2-sec sampling rate.}
\label{fig:histograms}}
\end{figure}

\subsection*{Path definitions} Paths were defined as the sequence of line segments walked by the termite individual every 0.5 seconds inside the Petri dish (Fig. \ref{fig:distances}). Waiting times are the time elapsed when the distance recorded was zero.
Both quantities, distances and waiting times, are heavily dependent on the sampling rate. Under-sampling occurs if the time interval between two samples misses turning points and oversampling happens when two video shots are taken from the same line segment. Avoiding both, under- and over-sampling, is a matter of prime importance. However, it is not always clear how to do it. From Fig. \ref{fig:histograms} it is obvious that different sampling rates would result in different conclusions if an histogram is the only choice made in order to measure potential scaling exponents, coming from hypothetical power-laws. Current trend of model fitting may even led toward selecting a model that excels the statistical criteria but has no phenomenological meaning. Here, we circumvent this,  completing a series of alternate data analysis to evince key aspects of the termite walking behaviour while, at the same time, its dynamical richness is exhibited.

\section*{Results}

\subsection*{MSD and anomalous difussion}

\begin{figure}[!h]
\centering
\includegraphics[width=80mm]{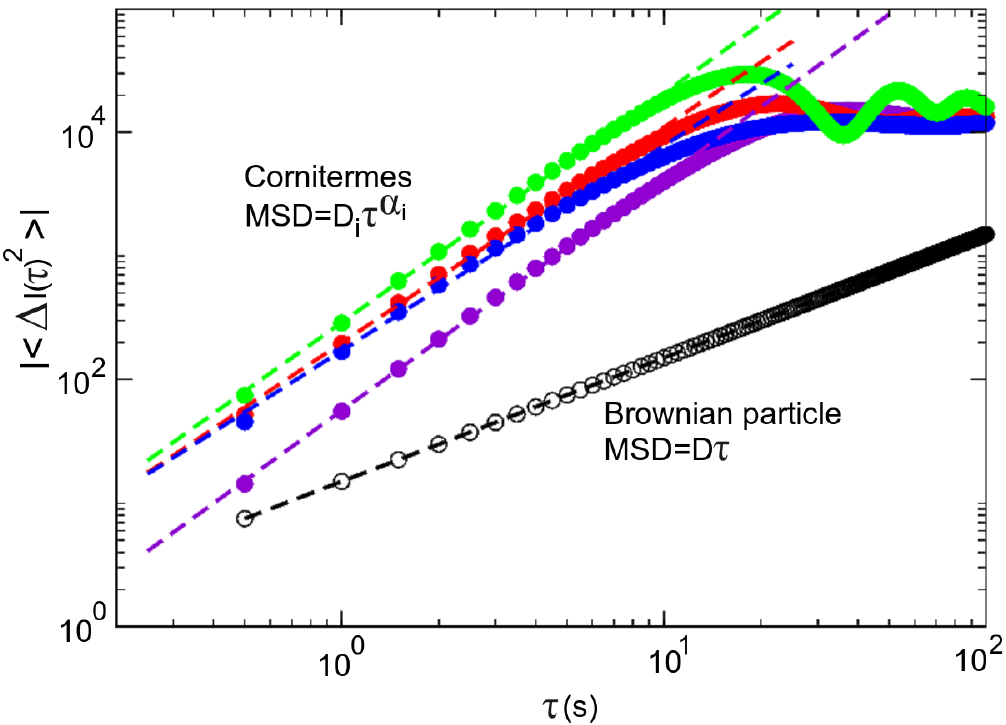}
\caption{{\small{ Anomalous diffusion.}\emph{Cornitermes cumulans} termites exhibit anomalous diffusion ($\alpha > 1$) in their walking patterns because the mean squared displacement grows faster than it does in the normal diffusion of a Brownian particle (black), where $\alpha=1$.  MSD superdiffusive scaling exponent values of four termite workers are $\alpha=1.90$ (purple), $\alpha=1.86$ (green), $\alpha=1.66$ (blue) and $\alpha=1.75$ (red). Notice that the termite MSD scaling separate away from a power-law at values of $\tau > 10$ and beyond, this is common and correspond to the typical diffusive behaviour of truncated motion in confined environments. $D_i$ is the diffusion coefficient of each individual termite}}.
\label{fig:msd}
\end{figure}

A widely-used method to evaluate the diffusive properties of mobile objects is the mean-square displacement (MSD) $\langle\Delta \vec{r}_{\tau}(t)^ 2 \rangle$, where $ \Delta \vec{r}(t)= [\vec{r}(t)-\vec{r}(t+\tau)]$ is the object displacement at time $t$ with a time delay $\tau$. Normal diffusion, as in the case of Brownian motion, occurs when the MSD increases linearly so that $\langle\Delta \vec{r}_{\tau}(t)^ 2 \rangle=2Dt$, where D is the diffusion coefficient of the object. Diffusive regimes where the MSD is not linear in time but proportional to $t^\alpha$ with $\alpha \neq 1$ are known as anomalous diffusion. Furthermore, when $\alpha$ is the diffusion exponent and $\alpha < 1$, the process is called subdiffusive, and if $\alpha > 1$, the process is called superdiffusive.

\begin{figure}[!h]
\centering
\includegraphics[width=90mm]{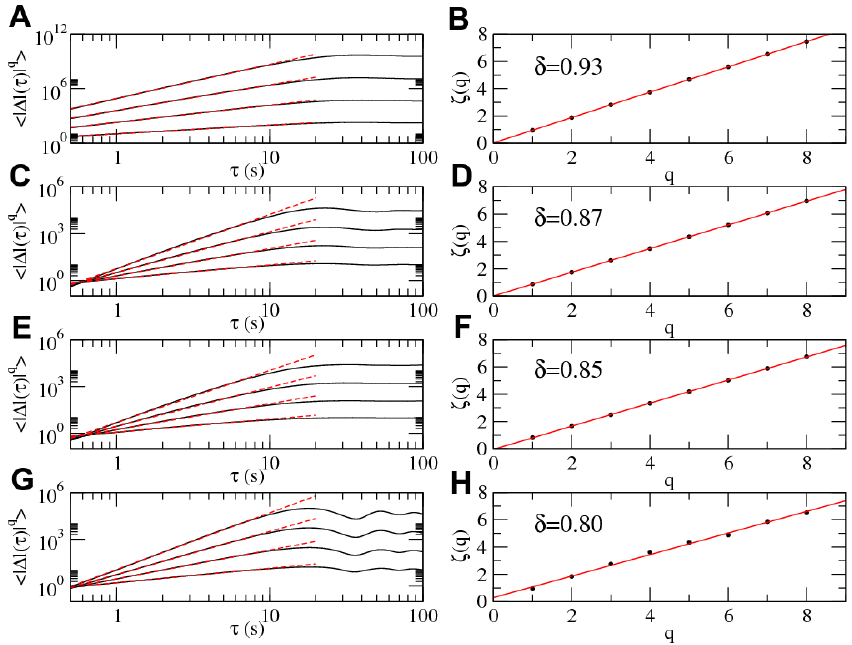}
\caption{{ \small{Kolmogorov $q$-functions.} Plots shown at the left column depict four examples of Kolmogorov $q$-functions and their power-law scaling (red lines). First eight values of the exponent $q$ were calculated but only four are shown for the sake of clarity. The column at the right depicts the linear scaling of $\zeta(q)$ (red lines), resulting in four $\delta$ slope values: 0.93, 0.87, 0.85 and 0.8. These correspond to L\'evy exponents of values $\mu=2.0$, $\mu=2.14$, $\mu=2.17$ and $\mu=2.25$, respectively. }}
\label{fig:q-functions}
\end{figure}

\emph{Cornitermes cumulans} diffusion in the arena can be estimated and characterized when the value of the MSD is calculated and when, for the sake of clarity, it is compared against a simulated Brownian particle of a known $\alpha=1$. It is easy to see, in Fig. \ref{fig:msd}, that while the Brownian particle has the expected scaling exponent of $\alpha = 1$, all termite's scaling exponents are $\alpha > 1$ and so their movements are properly superdiffusive. Termite step lengths and waiting times statistics (see Figs.\ref{fig:q-functions} and \ref{fig:waiting-times}) have broad distributions corresponding to superdiffusive (L\'evy) process where the mean squared displacement grows faster than it does in the normal diffusion of a Brownian particle \cite{klafler2005anomalous}. Phenomenologically, termite workers exhibit anomalous diffusion in their walking patterns because these individuals remain in motion without changing direction for a time that follows a scale-free distribution. These less-frequent long distance displacements are clearly visible in our experimental arenas (Fig. \ref{fig:tra}).

Physical systems where superdiffusion and L\'evy fligths were first identified together and  theoretically related to each other, have been studied in great detail up to now. These systems include the transport of passive particles in non-linear fluids where fluid motion may be retarded at the border of regions having different velocity vectors. Such regions cause ``sticking" that result in long-tailed distributions, both in the distances traveled and the waiting-times \cite{solomon1993observation}. For all practical purposes both superdiffusion and L\'evy fligths are two sides of the same coin, in such a way that a measure of one is related to a measure on the other. Superdiffusion and L\'evy fligths scaling exponents ($\alpha$ and $\mu$) are know to be theoretically related by a simple expression \cite{klafter1987stochastic,solomon1993observation}:

\begin{equation}
\label{relation1}
\mu = 4 - \alpha,
\end{equation}

\noindent that can be used to estimate  L\'evy fligths exponents from a MSD measure, as detailed below.

\subsection*{Structure functions and scaling}

 A well know method to characterize potential free-scale behaviour is due to Andrey Kolmogorov in the context of turbulent flows \cite{frisch1991global, yu2003structure, Arenas.Chorin.2006}. A generalization of the Kolmogorov method is known as structure functions of order $q$, being expressed as $\langle|\Delta l_{\tau}|^q \rangle\approx \tau^{\zeta(q)}$, \noindent where $l$ is the traveled distance between two ($x$,$y$) points in the space, $\Delta l$ is the distance increments, $\tau$ is a time lag that does the role of spatially coarse-graining the data set at different scales, so that scale-invariant observables could be spotted through the scaling of the exponent function $\zeta(q)$ and, finally, the angled brackets denote an average operator. Since the 1941 Kolmogorov seminal work on turbulence, the $q$-order structure functions have been used extensively in physics \cite{Benzi.etal.1993,Chechkin.Gonchar.2000,padoan2002structure,padoan2003structure,padoan2004structure,chapman2005scaling}, its use in biology for studying scale-free movement patterns in animals is relatively new and mostly applied to the analysis of invertebrates such as copepods \cite{Schmitt.Seuront.2001} and fruit flies \cite{Reynolds.Frye.2007}.

In order to study the free-scale patterns of termite movements under the Kolmogorov framework, we considered the trajectories yield by the walking of representative solitary individuals confined in a Petri dish arena where the values of the re-scaling lag $\tau$ was estimated from

\begin{equation}
\label{funciones2}
|\Delta l_{\tau}|=((x(t+\tau)-x(t))^{2}+((y(t+\tau)-y(t))^2)^{1/2}.
\end{equation}

Then the averages over all increments were calculated and elevated to the proper $q$ power, so that the $q$-structure function is calculated: $\langle|\Delta l_{\tau}|^q \rangle \approx \tau^{\zeta(q)}$
in such a way that $log(\langle|\Delta l_{\tau}|^q \rangle)\approx log(\tau^{\zeta(q)})\approx \zeta(q)log(\tau)$. Notice that the last expression is a linear equation with slope $\zeta(q)$ that can be easily evaluated. The last step needed under this procedure is to realize that $\zeta(q)=\delta q$ and that the value of the slope $\delta$ is related to the {L\'evy} exponent $\mu$ as $\delta=(\mu-1)^{-1}$. It is also important to remark that if $\zeta(q)=\frac{1}{2}q$, the process is normal-diffusive but if $\zeta(q)\neq \frac{1}{2}q$ then the process exhibits anomalous diffusion, with $\zeta(q) > \frac{1}{2}q$ being the superdiffusive regime (a condition fulfilled by all termites examined here).

 Equation \ref{funciones2} is used to estimate the structure functions $\langle|\Delta l_{\tau}|^q \rangle$ of our time series for $q$-orders $=\{1,2..,8\}$. Results are plotted in Fig. \ref{fig:q-functions} (left column) where linear scaling is seen on the log-log plot for values of $\tau < 10$, therefore a power-law scaling is evident. The values of these slops are, in turn, plotted in Fig. \ref{fig:q-functions} (right column) These plots evince the linear scaling behaviour of $\zeta(q)$ and so the value of the slopes $\delta=(\mu-1)^{-1}$, where $\mu$ is the L\'evy exponent.

L\'evy flight exponents in the termite superdiffusive walking estimated by the two previous methods are compared in Table \ref{table:levy}. Notice that the mean values for the $\mu$
exponents gives $\langle\mu_1\rangle$$\pm$sd = 2.20$\pm$0.11 (from MSD) and $\langle\mu_2\rangle$$\pm$sd = 2.16$\pm$0.04 (from $q$-functions) that are remarkable similar. 

\begin{table}[!ht]
\centering
\caption{\bf {Scaling exponents of four representative examples of termite walking.}
}
\begin{tabular}{|c|c|c|c|c|}
\hline
Series & $\alpha$ (MSD) & $\delta$  & $\mu_1=4-\alpha$ & $\mu_2=(1/\delta) +1 $ \\
\hline
Termite A          & 1.90 & 0.93 & 2.10 & 2.08\\
Termite B          & 1.88 & 0.87 &  2.12 & 2.15   \\
Termite C          & 1.75 & 0.85 & 2.25 & 2.18  \\
Termite D          & 1.66 & 0.80 & 2.34 & 2.25   \\
\hline
\end{tabular}
\begin {center}\small{The values of exponents $\alpha$ and $\mu_1$ are related by theoretical results \cite{solomon1993observation}, \\
and are also compatible with the results of a L\'evy exponent $\mu_2$ via \\
a Kolmogorov structure functions analysis.}
\end{center}
\label{table:levy}
\end{table}

\subsection*{Distance temporal fluctuations}

The self-similar features of the distances traveled by a termite walking can be visualized  by means of a Fast Fourier Transform that converts time fluctuations into a frequency fluctuations domain. This is useful for detecting temporal features such as periodicity or broad frequency spectra. The analysis is also very illustrative because it helps detecting self-similarity when the power spectrum scales as a power-law that, in turn, may hint the presence of colored noise, self-organization or critical dynamics. In addition, and for comparison purposes, it is possible to artificially generate a power-law time-series using well know algorithms in order to exhibit the resemblance of both power spectra. This procedure adds further support in characterizing the termite walking pattern as a fractal process in the traveled distance distributions.
 
\begin{figure}[!h]
\centering
\includegraphics[width=75mm]{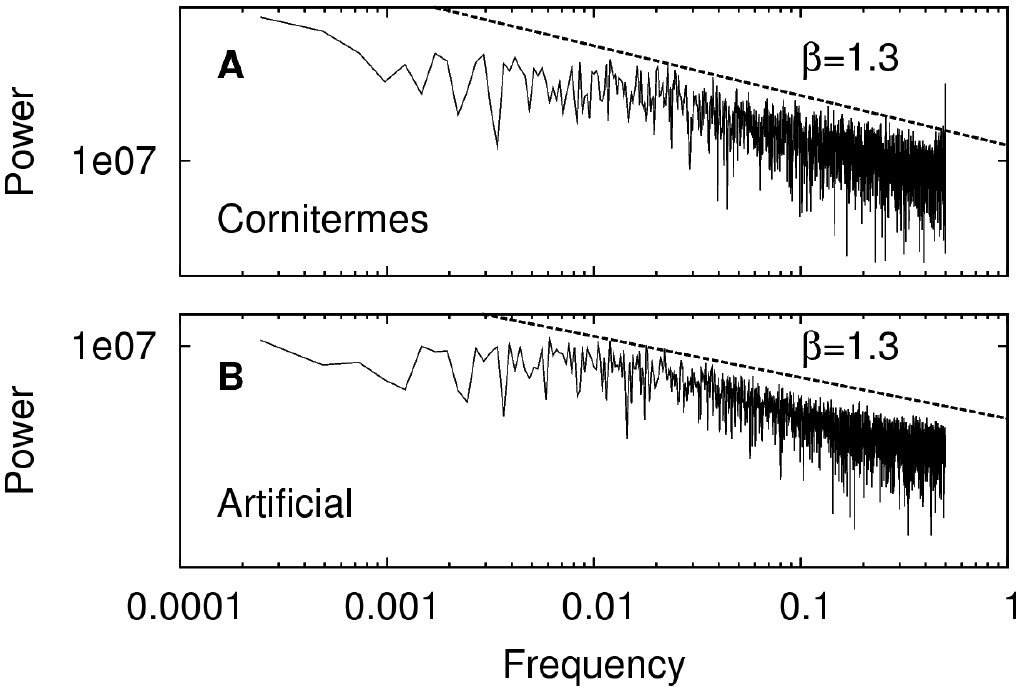}
\caption{\small{Power spectrum of a \emph{Cornitermes cumulans} termite walking time-series (A) and an artificially generated one (B). Both time series contained 4096 points and were transformed with a Fast Fourier Transform (FFT) algorithm.}}
\label{fig:FFT}
\end{figure}

For generating the artificial time-series, a transformation method between random probability distributions is preferred \cite{newman2005power}. The method can generate a random power-law-distributed real number
 $x = x_{min}(1 - r)^{(-1/(\mu-1))}$ in the range $x_{min} \leq x < \infty$ with exponent $\mu$, when feed with a random real number $r$ uniformly distributed in the range $0 \leq r < 1$. After the Fourier transform was applied  to a 4096 long segment of the termite walking, its power-law scaling $P(f)\propto f^{-\beta}$ was evident (Fig. \ref{fig:FFT}A), scaling with an exponent $\beta=1.3$. This information was used to generate a power-law with the same scaling exponent (Fig. \ref{fig:FFT}B). The resemblance between both spectra is remarkable, adding support for considering the termite walking pattern as a power-law relaxation process.

\begin{figure}[!h]
\centering
\includegraphics[width=110mm]{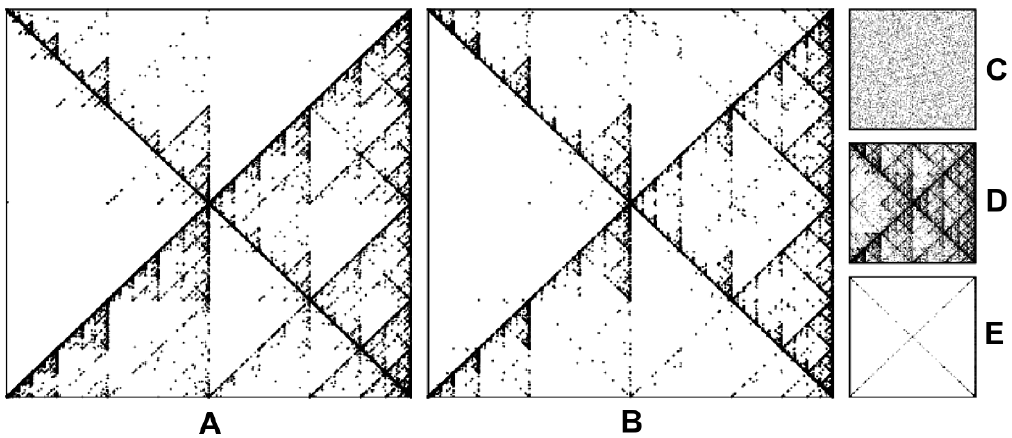}
\caption{\small{{IFS algorithm.} (A) A \emph{Cornitermes cumulans} termite walking time-series as seen with a IFS algorithm. Notice the subtle details of a self-similar structure. (B) An artificially correlated time-series generated with a  relaxation return map  $x_{t+1}=\eta x_t + (1-\eta)\epsilon_t$ where $\epsilon_t$ is a normally distributed random variable with zero mean and unit variance and $\eta$ is real valued parameter whose value determines the colour of the resulting time-series scaling. Colour in this context means the classification of a noise mode $1/f^{\beta}$, where $f$ is the frequency in a Fourier transformed space. $\beta$ is the scaling exponent and when $\beta=0$, the process is uncorrelated white noise (C), $\beta=1$ is correlated pink noise (D) and $\beta=2$ is a correlated brown noise (E). Termite walking (A) lies in between a pink (D) and brown noise (E) scaling, being compatible with the fact that the termite scaling exponent in the FFT is $\beta=1.3$. For details on how the IFS algorithm operates, see \cite{miramontes1998intrinsically}.}}
\label{fig:ifs}
\end{figure}

\subsection*{Fractal signatures and IFS}

L\'evy flights have spatial self-similarity due to their power-law nature. Because of this, time-series containing L\'evy-distributed random variables can be visualized to evince their subtle structural details, a pattern shared with time-series containing correlated coloured noise that can be generated artificially. An example of this is given in Fig. \ref{fig:ifs} where a fractal recurrence method known as Iterated Function System (IFS),  has been used to reveal the detailed self-similarity found in a termite walking.

IFS fractal reconstruction is a method widely used to visualize biological data such as in DNA sequences \cite{jeffrey1990chaos}, insect population dynamics \cite{miramontes1998intrinsically} or the temporal activity of social insects \cite{miramontes2001neural, Miramontes.DeSouza.2008}. Here the method is used for the first time to visualize the self-similar features of an animal walking while performing exploratory behaviour. Fig. \ref{fig:ifs} depicts one of the termite time-series. It reveals in full detail the self-similarity of it, on a set of fractal triangular structures. A similar picture arises when the IFS algorithm is used to visualize a time series of an artificial power-law distribution of a known FFT scaling exponent $\beta=1.3$. The foregoing result is sound evidence in favor of considering the termite walking pattern as a self-similar process with power-law distribution of step lengths.

\subsection*{Step correlations}

\begin{figure}[!h]
\centering
\includegraphics[width=70mm]{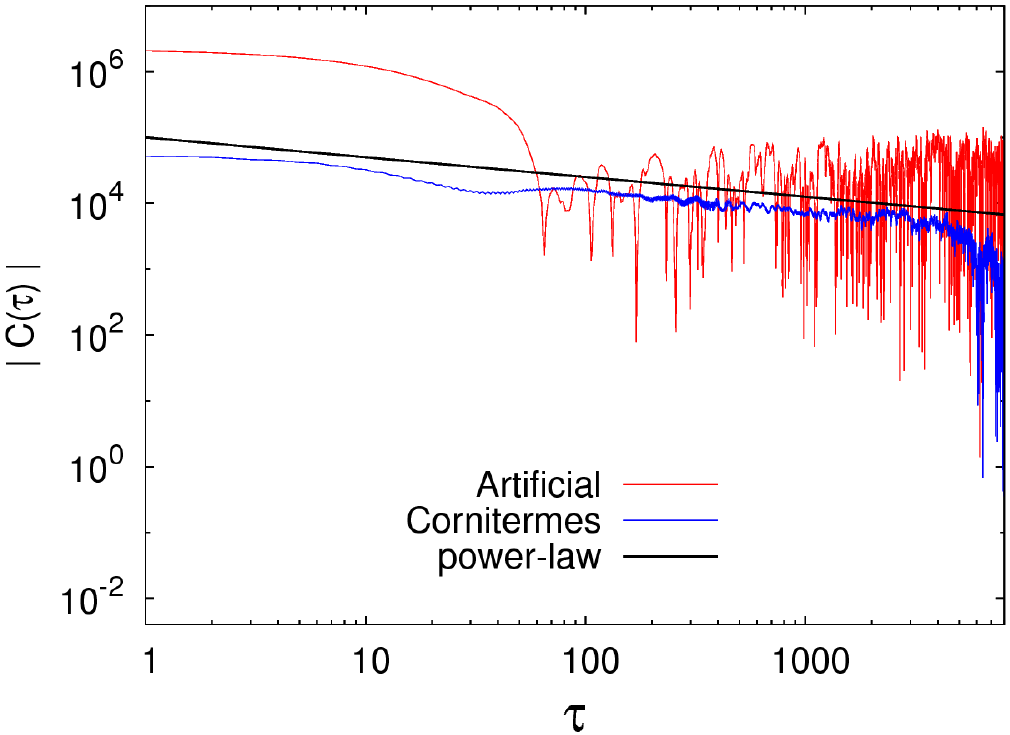}
\caption{\small{Long-range correlations. Termite walking exhibit power-law decaying long-range correlations as measured by a correlation function along the walk time-series (blue). An artificial correlated time-series, as explained in Fig. 8 was used also to compare a correlated decaying process (red). The black line is a power-law with a scaling exponent $\gamma=-0.3$.}}
\label{fig:correlation}
\end{figure}

Successive steps in a \emph{Cornitermes cumulans} termite motion are correlated to a certain extent, as  evident from the fact that these resemble correlated pink noise (Fig. \ref{fig:ifs}). In a termite walking, successive steps are no independent from each other with temporal dependencies that decrease along the trajectory. One method to identify and characterize such correlating distances is the correlation function (CF):

\begin{equation}
\centering
|C(\tau)|= \frac{1}{(N-\tau)\sigma^2} \sum_{t=1}^{N-\tau}[l(t)-\bar{\theta}][l(t+\tau)-\bar{\theta}], 
\end{equation}

\noindent where $N$ is the time-series length, $\tau$ is a time increment, $\sigma^2$ is the variance and $\bar{\theta}$ is the mean of the steps $l$. Correlation function is useful for finding temporal patterns in the time series such as periodicity or independence. In the case of white noise $|C(\tau)|=0$, so no correlation is present and the successive values are independent. We learn from Fig. \ref{fig:correlation} that this is not the case of termites. On the other hand, it is common that CF decays exponentially with rapid rates as $|C(\tau)|\propto e^{-\tau/\tau_s}$, in which case the time-series have correlations only among short distances $\tau_s$. Non-trivial correlation over long distances usually decay asymptotically as power-laws $|C(\tau)|\propto \tau^{-\gamma}$, with $0< \gamma <1$.  

\emph{Cornitermes cumulans} walking patterns have been found here to have long-range correlations decaying as power-laws. This means that a step length chosen by the termite at a particular time, will influence the length of future steps in a process akin to long-memory.  For years physicists have been studying systems that have power-law decaying CFs and have linked them to either systems which are far from equilibrium or very close to a critical point (e.g., a phase transition). We do not know the details of the locomotive physiological mechanism at play that generates a correlated termite walk but we can compare it to an artificially generated time series in order to gain more insights. Consider Fig. \ref{fig:correlation} where the same auto-correlated time-series explored before (Fig. \ref{fig:ifs}) is again analyzed. We now know that this artificial time-series has a power-law decaying CF but has originated by a multiplicative growth process. There are differences worth mentioning. Notice that the termite CF varies softly in contiguous values of $\tau$ while the artificial signal has a CF varying wildly in amplitude along of all its decay, evidencing its stochastic origin.

\subsection*{Waiting times}

\begin{figure}[!h]
\centering
\includegraphics[width=100mm]{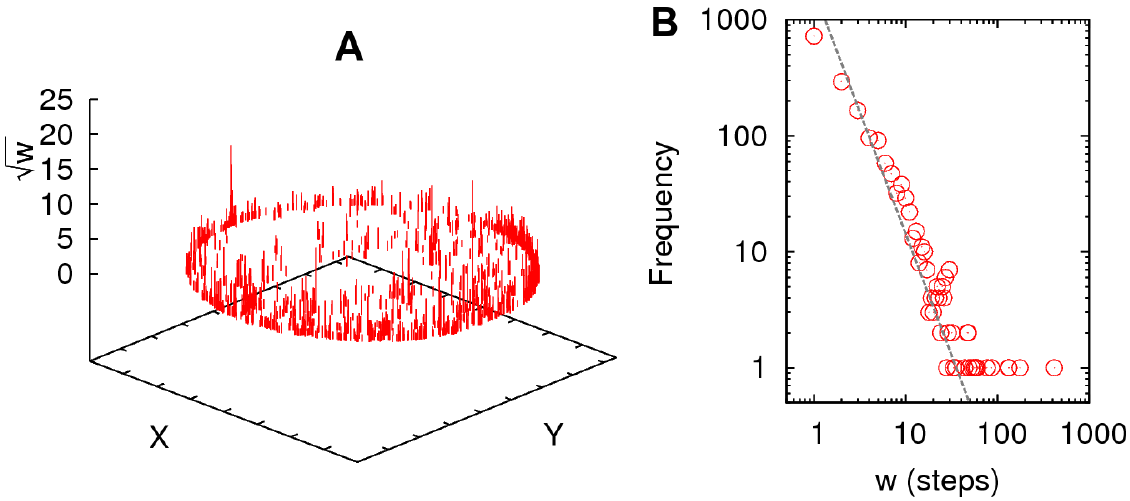}
\caption{\small{Waiting times. Waiting times are an ubiquitous pattern of animal movement behaviour and they may follow power-law scaling, as is the case of \emph{Cornitermes cumulans} workers when performing exploratory behaviour. The graph at the left (A) depicts a typical example of termite spatial distribution of accumulated waiting times over the circular arena (squared root axis for enhancing visualization). The plot in (B) is the waiting-time bouts histogram showing a power-law  with a scaling exponent value of $\lambda=2.64$ (straight line slope, calculated with a MLE procedure \cite{Clauset.etal.2009}.)}}
\label{fig:waiting-times}
\end{figure}

Foraging patterns of mobile individuals often exhibit resting periods known as waiting times. In an ecological context, it has been speculated that these may result from the encounter of places of interest that need time for being explored and exploited so that observed statistical properties of waiting times will reveal the distribution of resource patch sizes \cite{boyer2006scale}. In a behavioural approach waiting times may be originated in the fact that a mobile organism actually need periods of resting between activities. A number of foraging studies on the statistical properties of waiting times report that these behave as random-like burstiness that appear to be scale-free and so well-approximated by power-laws, meaning that short waiting periods are frequent while long are rare \cite{ramos2004levy}. It has also been recently speculated that the power-law scaling of waiting times is a behavioural ``rule of thumb'' evolved to optimize move-wait decisions in unpredictable environments \cite{wearmouth2014scaling}. Remarkably these waiting times resemble the scaling laws found in the study of superdiffusive passive trackers moving in non-linear flows and where these waiting times are a well-known indivisible aspect of anomalous diffusion \cite{solomon1993observation}. In fact, let's return to equation \ref{relation1} where a relationship between the values of the MSD and L\'evy scaling exponents was established. As a matter of fact, equation \ref{relation1} holds best if $\lambda>2$ \cite{zumofen1995laminar, ramos2004levy}. Since the equation actually holds in our case because the estimated values of $\mu_1$ and $\mu_2$ are consistent (Table 1), we expect that the termite waiting times will be described as power-laws $w(t) \propto t^{-\lambda}$ with $\lambda>2.0.$ We show in Fig. \ref{fig:waiting-times} that \emph{Cornitermes cumulans} worker termites indeed exhibit power-law waiting times when performing exploratory behaviour in navigational clueless environment, where no food or nestmates are present. Waiting times were calculated from the termite walking time series when the isolated termite workers did not changed position. Before walking again a termite may stop and wait for as long as 0.5 seconds to as much as 200 seconds. The log-log plot of the waiting time data, for the example shown, resulted in a negative power-law function with an exponent $\lambda=2.64$; other measured  values were 2.28, 2.52 and 2.96, with $\langle\lambda\rangle$$\pm$sd = 2.60$\pm$0.28.

\subsection*{Turning angle distribution}

\begin{figure}[!h]
\centering
\includegraphics[width=90mm]{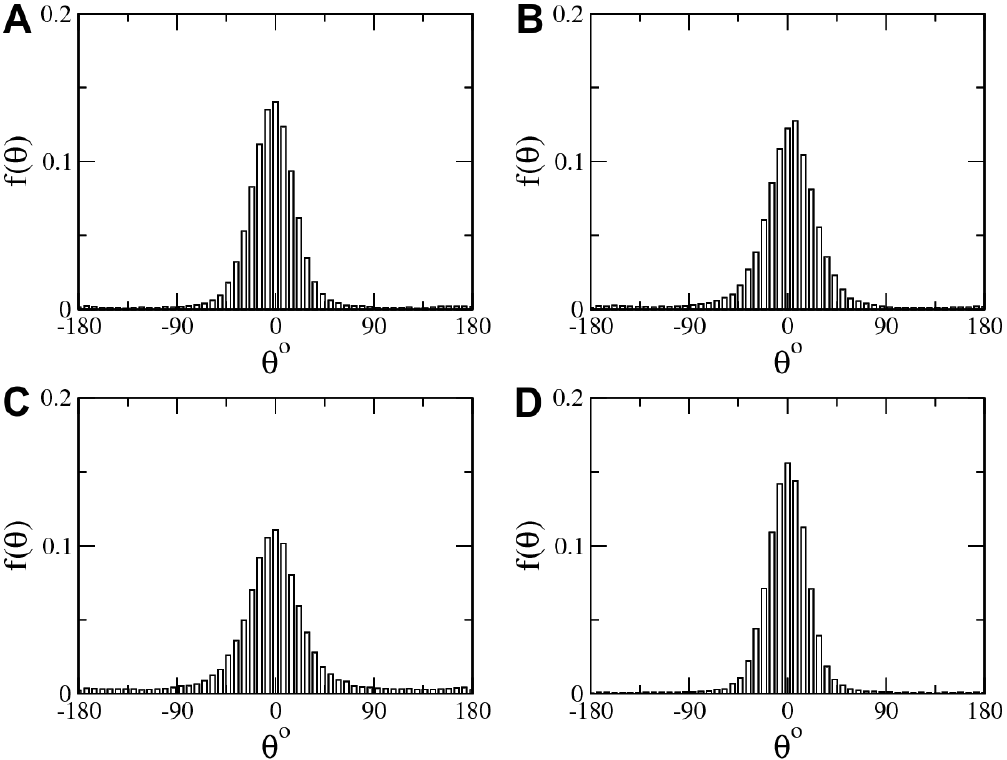}
\caption{\small{ Turning angle distribution. Turning angle distribution in termite walking. Four examples are depicted exhibiting a bell shaped distribution centered at 0 degrees. No preferential angles were identified apart from the persistence of moving forwards.}}
\label{fig:angles}
\end{figure}

Inside the entangled network of tunnels, termites follow well defined routes for traveling between different places of interest, from the foraging areas, to the royal chamber, etc. Here we are interested in exploring to what extent there is persistence in the direction of consecutive steps in termites that would reveal, for example, if termites move exploring space in preferential angles or not. This is  because it has been speculated that they may be doing movements with angles of about 40--60 degrees as an optimal searching strategy that may be related to the  branching angle of termite tunnels \cite{Bardunias.Su.2009}. When exploring the distribution of turning angles between successive steps, it is found here that \emph{Cornitermes cumulans} workers in open spaces do not show evidence of pre-wired preferential angles. There is a directional trend to move forwards but the resulting probability distribution is a uniform bell shaped curve centered in 0 degrees, as seen in Fig. \ref{fig:angles}. The absence of preferential angles of around 40--60 degrees may imply that the geometry of the tunnels may be either the outcome of social interactions or a response of termites to terrain clues, or a combination of both.

\section*{Discussion}

The study of animal movements is of prime importance for understanding ecological and behavioural traits of individual displacements, needed to the efficient use of space. These movements in a social context are extremely important because these regulate the rate of interactions that are the basis for building self-organization global patterns of behaviour \cite{Miramontes.DeSouza.2008} which are, ultimately, essential to cooperation phenomena and social stasis. Social individuals exchange information related to both the status of individuals and the social group and this information flows between individuals who move and encounter each other.

Our results show that free termite displacement, in space and time, present self-similar patterns highly consistent with anomalous diffusion. In social animals interacting nestmates are valuable resources to look for, and that is achieved by moving about in an efficient manner. Isolated termite displacement patterns revealed in this study, add support for the hypothesis that animals adopt L\'evy flights because these confer an advantage in terms of higher fitness resulting from greater efficiency in finding nestmates to whom socially interact. The current trend that highlights statistical model fitting has perhaps put to much attention in fitting methodologies, which are important of course, but other qualitative and quantitative aspects of the phenomena should not remain overlooked, specially the biological phenomenology. Superdiffusion, L\'evy flights and waiting times are quantitatively interconnected in ways that must be explored further. It is intriguing that passive trackers in fluid dynamics do behave qualitatively and quantitatively the same way that termites do. Superdiffusion is a movement pattern observed in a wide range of organisms thought to be related to optimal search strategies. Here we show that termites perform superdiffusive displacements, suggesting that this movement pattern is important for developing social interactions upon which self-organization at the colony level is built. Our results suggest that scale-free superdiffusion may have played an important role in the evolution of societies and cooperation because it may confer an advantage in terms of fitness to individuals that profit from efficient social encounter rates, that translates into efficient information transfer.

What is remarkable in our study, is the fact that assayed termites exhibited superdiffusion and L\'evy flights even in the absence of external stimuli or clues: the arenas contained no food nor other nestmates. It seems, therefore, that such a movement pattern is indeed hardwired in termite ``instincts''. Or recalling a long forgotten idea in termitology, termites orient themselves concatenating external stimuli with some form of internal momentum \cite{Jander.Daumer.1974}. The important connotation arising from such a result is that termite diffusion can proceed independently of a reactive phasis, as expected for a walker with limited (if any) cognitive  abilities \cite{Viswanathan.etal.2011}. More specifically, blind termites seem to be equipped with strategies which secure them to find their targets (food or nestmates) even in environments where these are scarce or criptically located.


\section*{Acknowledgments}
OM thanks DGAPA-PAPIIT Grant IN101712 and the Brazilian Ci\^encia Sem Fronteiras program (CSF-CAPES)  0148/2012. ODS is supported by CNPq-Brasil, fellowship 305736/2013-2. We very much appreciate valuable comments, suggestions and support from D. Boyer, D. Lemus, A. Chopps and C. Corona.


\bibliographystyle{IEEEtran-esp}
\bibliography{biblio}

\end{document}